\documentclass{article}
\usepackage{graphicx} 
\usepackage{url} 
\title{Efficient Search in Graph Edit Distance: Metric Search Trees vs. Brute Force Verification}
\author{Wenqi Marshall Guo \\
wg25r@student.ubc.ca
\and Jeffrey Uhlmann\\
uhlmannj@missouri.edu
}
\date{March 2024}

\begin{document}

\maketitle

Technical Report for Multidisciplinary Undergraduate Research Conference (MURC) 2024 \footnote{For more information, please access \url{https://github.com/weathon/GED-CMT}}.

\section{Introduction}

Graph similarity search (GSS) plays a key role in domains such as molecular and protein similarity search[1]. Specifically, Graph Edit Distance (GED)--the measure of similarity based on the steps required to edit one graph into another--is one of the most commonly used metrics in GSS. However, the computation of exact GED traditionally requires exponential time, which can slow down the search process considerably[1].

Previous researchers have proposed various methods to speed up the GED-based GSS process. One of them is a brute-force-based method that first uses lower bounds to filter the graphs in the database (only the graphs with lower bounds less than the threshold will be kept) and then verification in which the computation of exact GED is replaced by a procedure to verify if the GED of two graphs is lower than a given threshold[1].

Another promising approach involves augmenting similarity searches with metric search trees, such as the Cascading Metric Tree (CMT)[3]. CMT has been shown to enhance similarity searches using different metrics, like Euclidean distance[3] and Kendall-Tau distances[2]. In this study, we aim to investigate whether adapting the CMT for use with GED using upper-and-lower bounds (UBLB) can outperform the simple brute force verification relayed in [1]. Our results suggest that, contrary to our initial hypothesis, CMT does not routinely outperform brute force verification strategy in most contexts.

\section{Methodology}

As mentioned previously, exact GED is extremely hard to compute, this leads to the use of UBLB in the querying process of the algorithm. However, since the tree will be pre-built only once, to enable more effective pruning, the exact distance calculation is still employed during the tree's construction phase.

Our query algorithm could be described as follows:

\begin{itemize}
    \item Two empty answer sets will be prepared, 'confirmed set' and 'suspected set'
    \item The UBLB information is used to prune the search tree with a similar process in [3].
    \item If the upper bonds of the largest distance in a subtree are less than the threshold, all the graphs of the subtree will be added to the confirmed set.
    \item When it reaches the leaf nodes:

\begin{itemize}
        \item If the upper bounds are lower than the threshold, the corresponding graph is added to the confirmed set.
        \item If the lower bounds are lower than the threshold, the graph is added to the suspected set.
\end{itemize}

    \item A brute force verification step then examines each member of the suspected set to filter out any false positives from the UBLB search.
\end{itemize}
We tested our implementation on Euclidean distance to debug and ensure it is working as intended. Then, the algorithm was tested on graph data derived from PubChem and the performance of CMT was compared to that of brute force verification.

\section{Results and Discussion}
\includegraphics[width=0.5\textwidth]{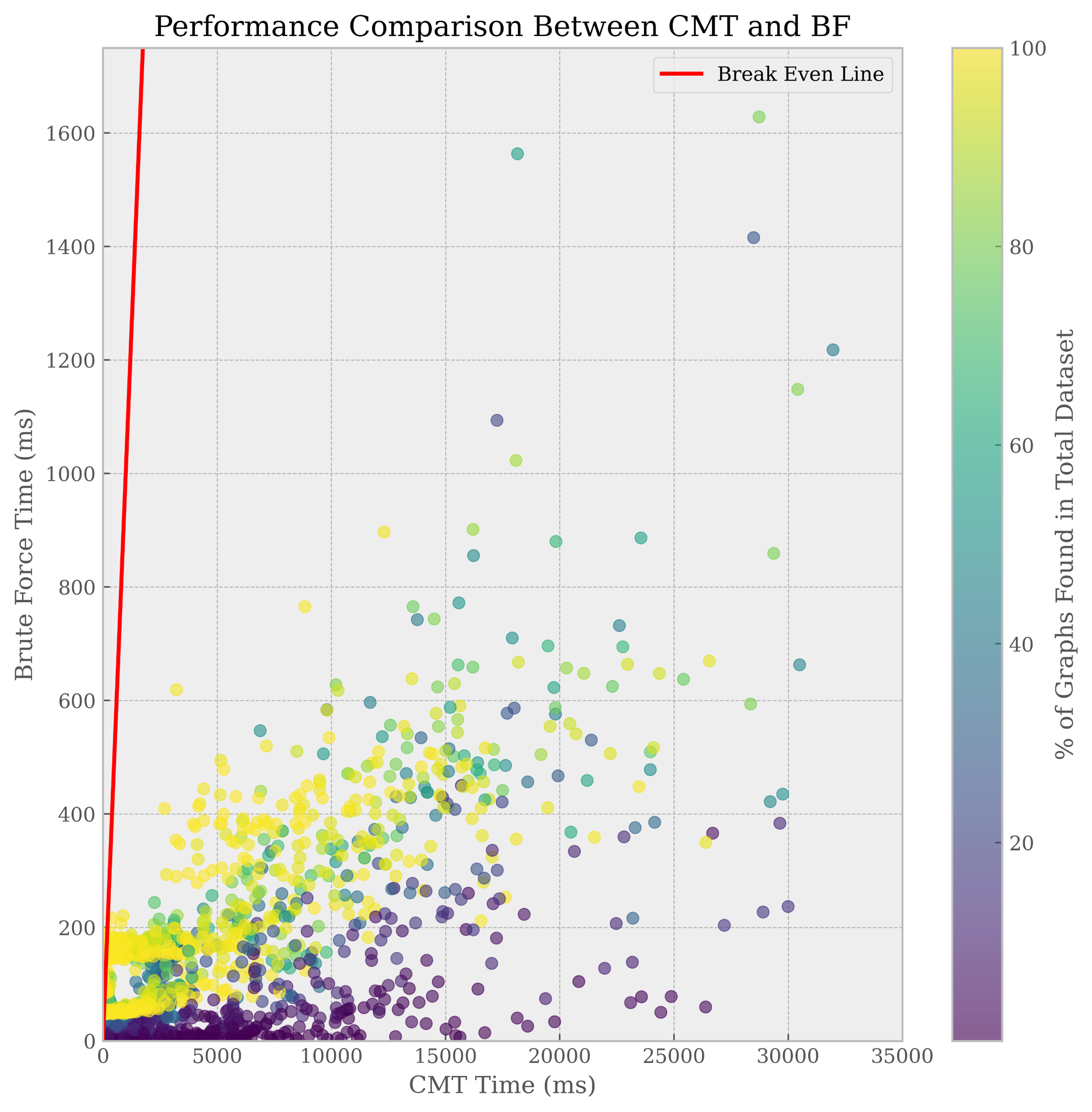}
\includegraphics[width=0.5\textwidth]{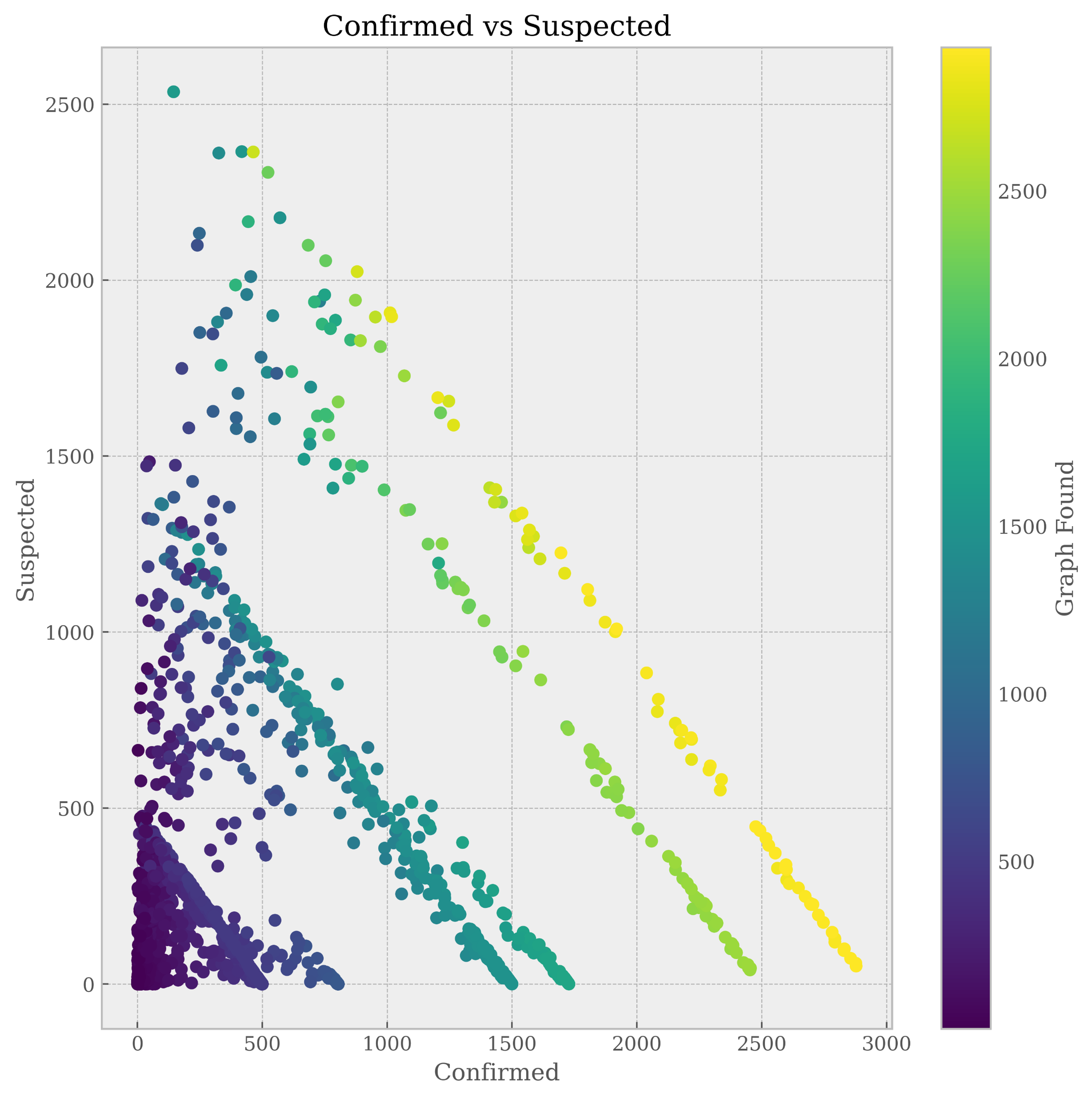}
\includegraphics[width=0.5\textwidth]{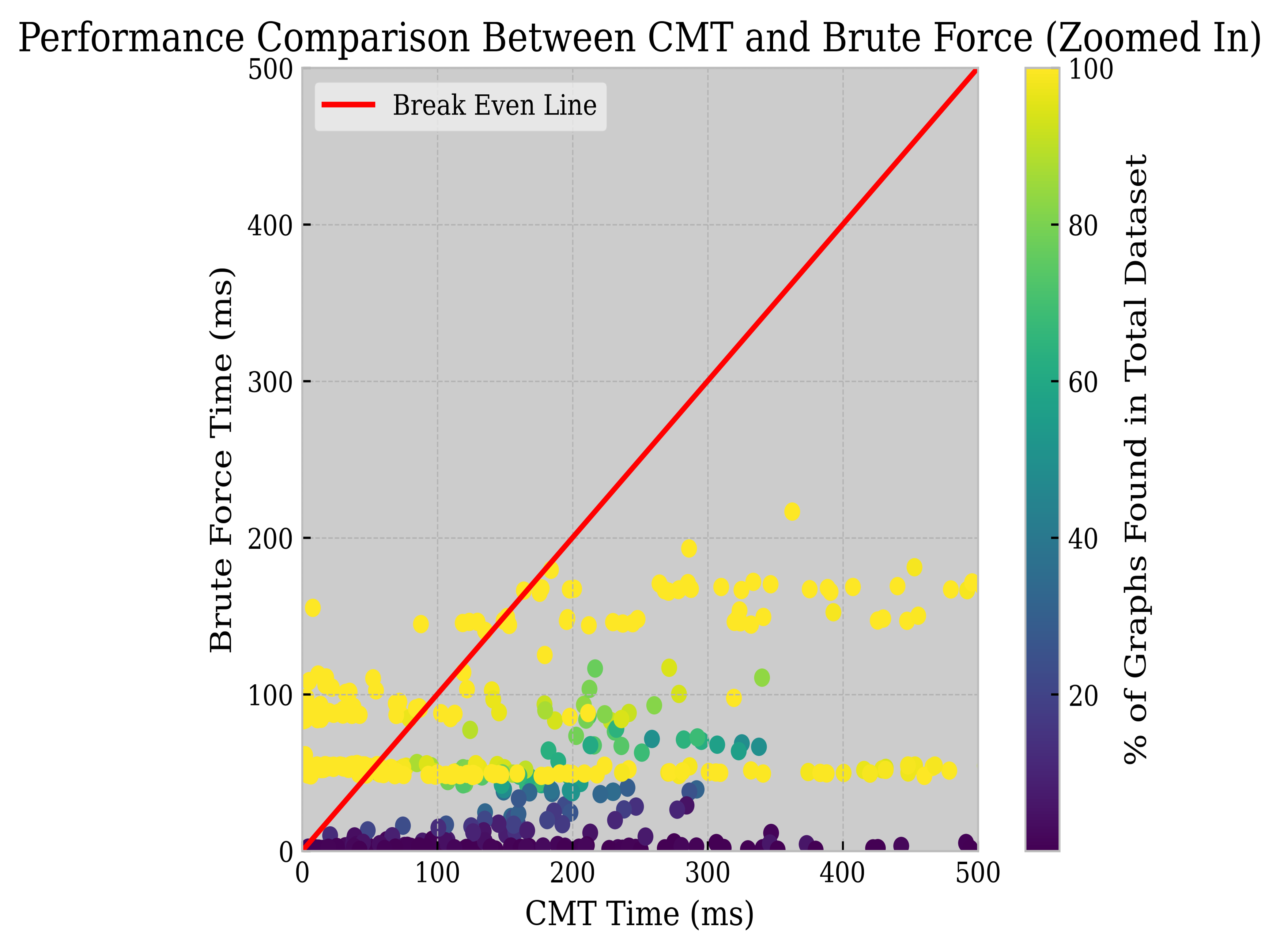}
\includegraphics[width=0.5\textwidth]{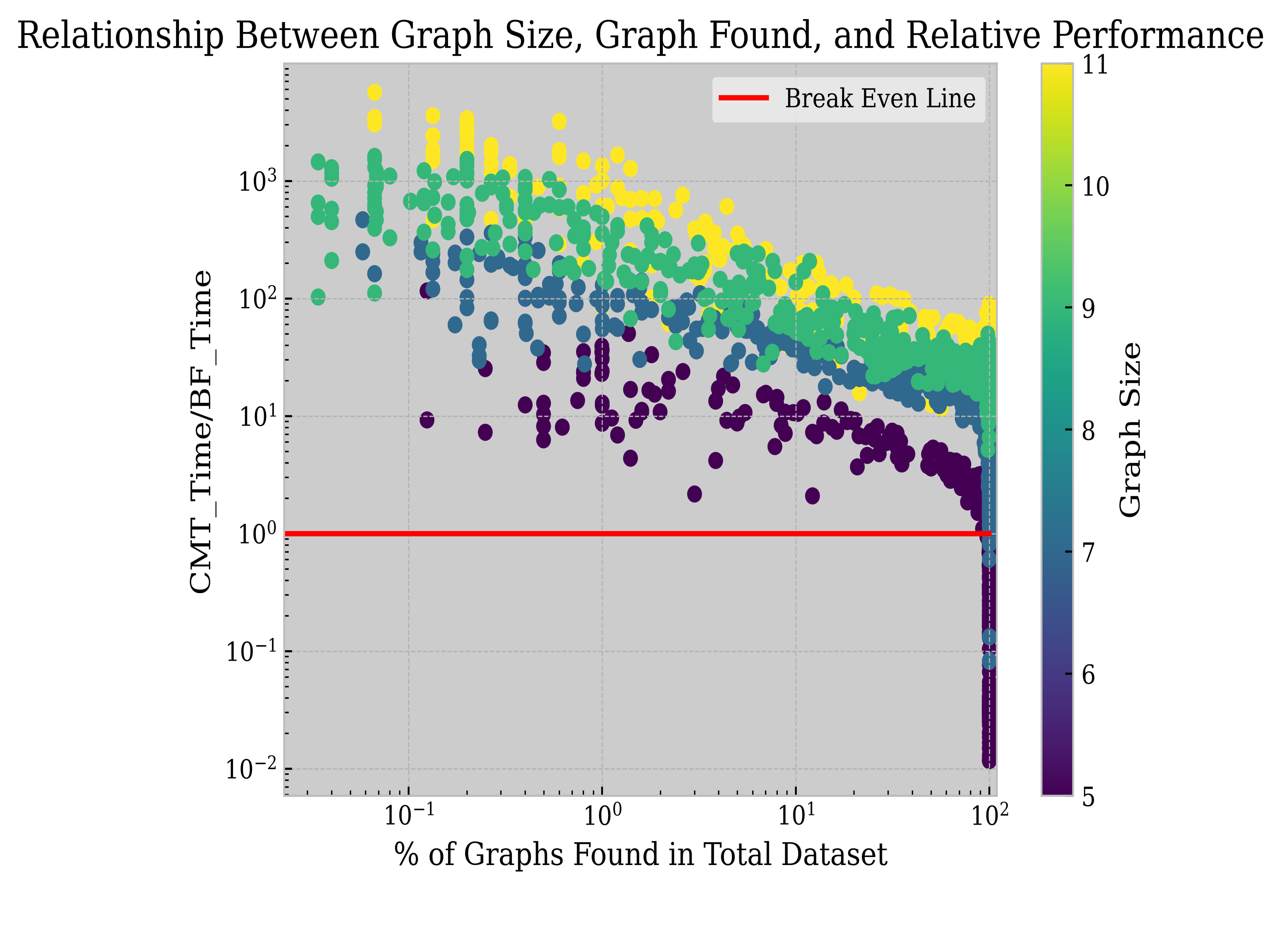}
\includegraphics[width=0.5\textwidth]{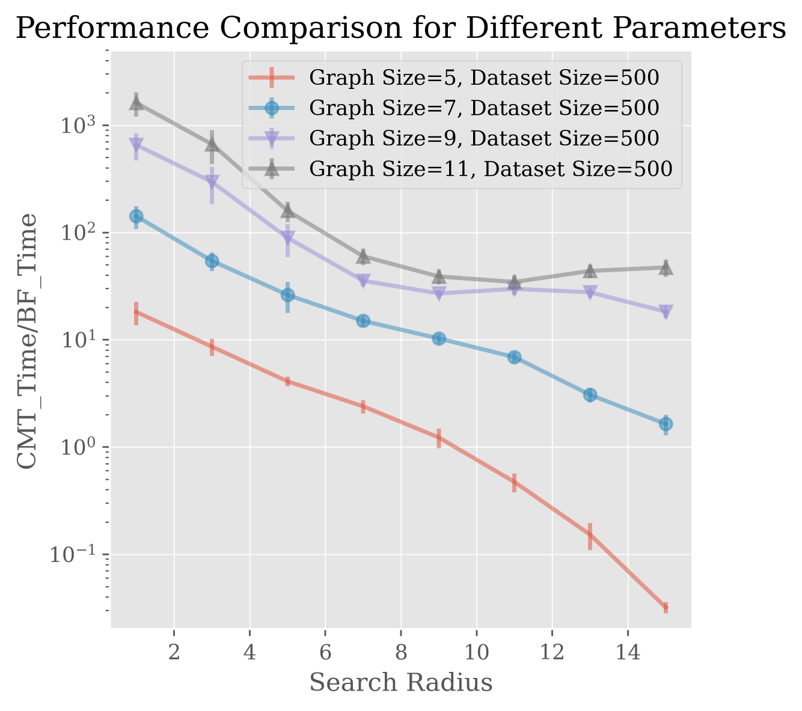}
\includegraphics[width=0.5\textwidth]{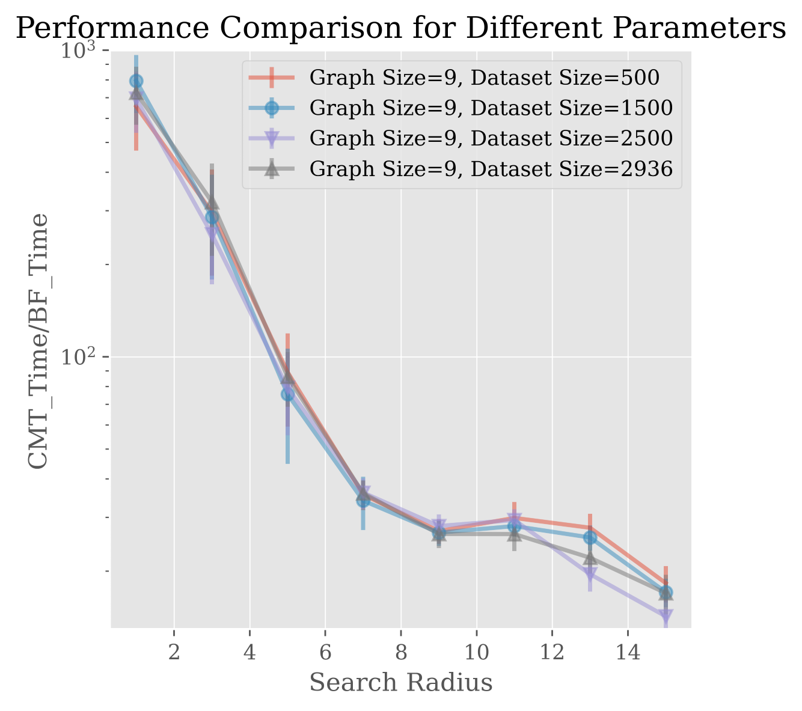}
   Results indicate that CMT does not routinely outperform brute force verification. In fact, in many cases, CMT significantly lags behind brute force in terms of speed. One possible explanation for this could be even for UBLB of the GED, the computation is still expensive. On the other hand, the verification could be done very fast with a small query radius (which is usually the case for practical applications). This surprising finding underlines the need for continuous innovation in methodologies to speed up GED computation.

\section{Future Work and Limitations}

Our study is a preliminary investigation of the relative performance between tree search strategies like CMT and brute force verification and it has several limitations which indicate possible directions for future work. These include:

There are different hyper-parameters that will affect the result such as the iterations of the upper and lower bounds algorithm. Different hyper-parameters might yield different results.

We only tested the algorithm in a very limited search space. The dataset size could largely limit the performance of CMT.

We only used the graph data from PubChem, there might be a bias in the dataset and the result could be different on other datasets.

The structure of an efficient search tree typically leads to better relative performance with larger datasets, a trend partially confirmed by our study. At the same time, our research also suggests that smaller graphs may deliver better relative performance. However, this is a “blank space” in our dataset: we do not have many small molecules. Examining this space might be valuable to find potential space where CMT is more efficient.

We did not optimize our implementation of the CMT, by optimizing it the performance could be improved.

\section{Conclusion}

This study investigated the performance of Cascading Metric Trees in GED computation for graph similarity search in comparison to brute force verification. Surprisingly, our results showed that despite its proven efficacy in other metrics, CMT did not routinely outperform brute force verification in computing GED. These findings highlight the need for continued exploration of methods for speeding up GED computation and verification, with a particular focus on optimizing the computation of GED upper and lower bounds and investigating the performance of these methods on a wider range of datasets and search spaces.

\section{ACKNOWLEDGEMENT}

Part of the computation for this work was performed on the high-performance computing infrastructure provided by Research Support Solutions and in part by the National Science Foundation under grant number CNS-1429294 at the University of Missouri, Columbia MO. DOI: \url{https://doi.org/10.32469/10355/69802}{https://doi.org/10.32469/10355/69802}

OpenAI ChatGPT is used as an aid in this project such as brainstorming, language refining, etc. This tech report is produced by ChatGPT based on an author-written outline and proofread and largely edited by the author.

\section{References}

[1] L. Chang, X. Feng, X. Lin, L. Qin, W. Zhang, and D. Ouyang, ‘Speeding up ged verification for graph similarity search’, in 2020 IEEE 36th International Conference on Data Engineering (ICDE), Apr. 2020, pp. 793–804. doi: 10.1109/ICDE48307.2020.00074.

[2] W. Guo and J. Uhlmann, ‘Metric search for rank list compatibility matching with applications’. arXiv, Aug. 10, 2023. doi: 10.48550/arXiv.2303.11174.

[3] J. Uhlmann and M. R. Zuniga, ‘The cascading metric tree’. arXiv, Dec. 20, 2021. doi: 10.48550/arXiv.2112.10900.

[4] D. B. Blumenthal, S. Bougleux, J. Gamper, and L. Brun, ‘Gedlib: a c++ library for graph edit distance computation’, in Graph-Based Representations in Pattern Recognition, D. Conte, J.-Y. Ramel, and P. Foggia, Eds., in Lecture Notes in Computer Science. Cham: Springer International Publishing, 2019, pp. 14–24. doi: 10.1007/978-3-030-20081-7\_2.

[5] Z. Abu-Aisheh, R. Raveaux, and J.-Y. Ramel, ‘A graph database repository and performance evaluation metrics for graph edit distance’, in Graph-Based Representations in Pattern Recognition, C.-L. Liu, B. Luo, W. G. Kropatsch, and J. Cheng, Eds., in Lecture Notes in Computer Science. Cham: Springer International Publishing, 2015, pp. 138–147. doi: 10.1007/978-3-319-18224-7\_14.

[6] K. Riesen, A. Fischer, and H. Bunke, ‘Computing upper and lower bounds of graph edit distance in cubic time’, in Artificial Neural Networks in Pattern Recognition, N. El Gayar, F. Schwenker, and C. Suen, Eds., Cham: Springer International Publishing, 2014, pp. 129–140. doi: 10.1007/978-3-319-11656-3\_12.

\end{document}